\documentclass[conference]{IEEEtran}
\IEEEoverridecommandlockouts
\usepackage{cite}
\usepackage{amsmath,amssymb,amsfonts}
\usepackage{algorithmic}
\usepackage{graphicx}
\usepackage{textcomp}
\usepackage{xcolor}
\usepackage{caption}
\usepackage{subcaption}
\usepackage{hyperref}
\usepackage{stfloats}

\def\BibTeX{{\rm B\kern-.05em{\sc i\kern-.025em b}\kern-.08em
    T\kern-.1667em\lower.7ex\hbox{E}\kern-.125emX}}

\newcommand{\siva}[1]{}
\newcommand{\vasu}[1]{}
\newcommand{\steven}[1]{}
\newcommand{\michal}[1]{}

\begin{document}

\title{ALBERTA: ALgorithm-Based Error Resilience in Transformer Architectures \\
}


\author{
Haoxuan Liu, Vasu Singh, Micha\l{} Filipiuk, and Siva Kumar Sastry Hari \\ 
NVIDIA \\
\{steveliu,vasus,mfilipiuk,shari\}@nvidia.com
}
\maketitle

\begin{abstract}
Vision Transformers are being increasingly deployed in safety-critical applications that demand high reliability. It is crucial to ensure the correctness of their execution in spite of potential errors such as transient hardware errors. We propose a novel algorithm-based resilience framework called ALBERTA that allows us to perform end-to-end resilience analysis and protection of transformer-based architectures. First, our work develops an efficient process of computing and ranking the resilience of transformers layers. We find that due to the large size of transformer models, applying traditional network redundancy to a subset of the most vulnerable layers provides high error coverage albeit with impractically high overhead. 
We address this shortcoming by providing a software-directed, checksum-based error detection technique aimed at protecting the most vulnerable general matrix multiply (GEMM) layers in the transformer models that use either floating-point or integer arithmetic. Results show that our approach achieves over 99\% coverage for errors that result in a mismatch with less than 0.2\% and 0.01\% computation and memory overheads, respectively. Lastly, we present the applicability of our framework in various modern GPU architectures under different numerical precisions. We introduce an efficient self-correction mechanism for resolving erroneous detection with an average of less than 2\% overhead per error.
\end{abstract}

\pagestyle{plain}
\section{Introduction}

Deep learning has become the key technology in several safety-critical fields like
autonomous vehicles (AVs)~\cite{av-survey}, medical image analysis~\cite{medical-image-analysis-DL}, and drug design~\cite{drug-design}. While convolutional neural networks (CNNs) were the predominant architecture of choice for deep learning in the last decade in computer vision tasks, the focus is rapidly shifting to transformer-based networks. Following recent breakthroughs in self-attention and model training optimizations, vision transformers (ViTs) have demonstrated their effectiveness in adapting to a wide variety of vision problems ranging from image classification~\cite{dosovitskiy2020image}, object detection~\cite{swin-transformer}, image generation~\cite{styleswin}, and semantic segmentation~\cite{khan2022transformers}. 

In most ViT-based architectures, latent dependencies between embedded image tokens are learned in parallel via the inter-/intra-token calculations computed by the self-attention modules. Processing an image in such a fashion remains computationally expensive today, primarily due to the high computation and memory cost of large matrix operations. 

It is important to understand the effect of transient hardware errors on various components of the transformer architecture. Such understanding can provide insights to develop resilient transformers without high computational overhead for safety-critical applications.

Prior work on resilience in deep neural networks utilize 
different techniques like range-based bounds checking~\cite{Chen2021}, redundant layer
computation~\cite{mahmoud2021optimizing}, and algorithm-based fault tolerance (ABFT) techniques~\cite{huang1984algorithm}. 
These methods are either not sufficient or pose significant overhead for vision transformers. 
We investigate the vulnerability of individual layers of
transformers and whether we can provide high resilience in 
vision transformer models with minimal overhead. 

In this paper, we present a framework called ALBERTA (ALgorithm-Based Error Resilience in Transformer Architectures) for end-to-end resilience analysis and enhancement framework for vision transformers. While ALBERTA can be employed using any deep learning framework, we prototype it using PyTorch~\cite{pytorch}.
To the best of our knowledge, this is the first paper to perform a detailed resilience analysis of vision transformer models (an important class of perception models) using floating-point and integer data types, provide novel insights into the vulnerable components, and propose a low-cost protection technique to significantly increase its resilience with ultra-low overheads. 
%
The following are the main contributions of this paper:

\begin{itemize}
\item 
We implement the vulnerability analysis in ALBERTA using an error injector module in PyTorch that allows the user to import custom datasets and select desired injection modes for transformer model analysis. The set of injection locations supported by the injector module include layer input, output, and weights, while the set of enabled injection types include single bit flip for integers, single bit in the exponent or mantissa for floating-point values, and replacement with random or user-specified value.     

\item We investigate the vulnerability of 
different layers in a transformer model and gather interesting insights: (1) Our chosen transformer architecture exhibits a significant jump in resilience  from the third to eighth multi-head attention blocks (out of twelve blocks), which is a counter-intuitive pattern compared to the findings on CNNs that earlier layers are resilient~\cite{Li2017, mahmoud2021optimizing}. 
(2) Injections in the prediction head of the transformer result in the highest probability of mismatch in the network output. 
(3) Selective layer duplication provides coverage of about 90\% at an overhead of more than
30\%.

\item The error-resilience algorithm in ALBERTA is also implemented as a Pytorch module. We implement optimized checksum computation for floating-point (FP) based models. It is well-known that such checksums cannot be bitwise precise due to FP non-associativity. So, the solutions have to rely on conservative thresholds for error checking, which can compromise error coverage. We present a methodology that derives the thresholds such that high error coverage is maintained. 

\item Our proposed error-resilience algorithm achieves over 99\% coverage of all injected errors that result in network misclassification at \textless0.2\% computation overhead and \textless0.01\% memory overhead. While false detections are rare, we introduce an efficient self-correction mechanism with an average overhead of \textless2\% to resolve each erroneous detection.

\end{itemize}

The paper is organized as follows.
Section~\ref{sec:problem_statement} formalizes our problem statement and  related work.
Section~\ref{sec:alberta} describes the ALBERTA framework and Section~\ref{sec:results} presents the results we 
obtain on different vision transformers using ALBERTA.
Section~\ref{sec:conclusions} concludes the paper and discusses further directions for future work.

\section{Problem Statement}
\label{sec:problem_statement}


%


\subsection{Background}
\label{sec:background}
\noindent {\bf Vision Transformers.}
Transformer-based models~\cite{vaswani2017} first gathered attention in the context of natural language processing. After their huge success, they started to gain traction in different areas~\cite{audio-transformer, timeseries-transformer}, computer vision including. 
The Vision Transformer (ViT) was one of the first transformer models for vision: it is pretrained (supervised) on the ImageNet-21k dataset at a resolution of 224x224 pixels, and later finetuned on ImageNet~\cite{imagenet} at the same resolution~\cite{dosovitskiy2020image}. Images are presented to the model as a sequence of 16x16 fixed-size patches that are linearly and positionally embedded. For image classification tasks, the model also contains an extra classification token that will eventually be passed through a final linear layer in order to produce the network result.
Studies have demonstrated that transformers do not generalize well when trained on insufficient amounts of data, and the training of these models involved extensive computing resources~\cite{touvron2021training}. As a result, DeiT was created as a more efficiently trained transformer that utilizes a novel distillation procedure. Specifically, the process involves a distillation token aimed at reproducing the label estimated by a teacher model (often a pretrained ConvNet). 

\begin{figure}[t]
    \setlength{\belowcaptionskip}{-5pt}
    \setlength{\abovecaptionskip}{1mm}
    \centering
    \includegraphics[width=0.3\textwidth, clip]{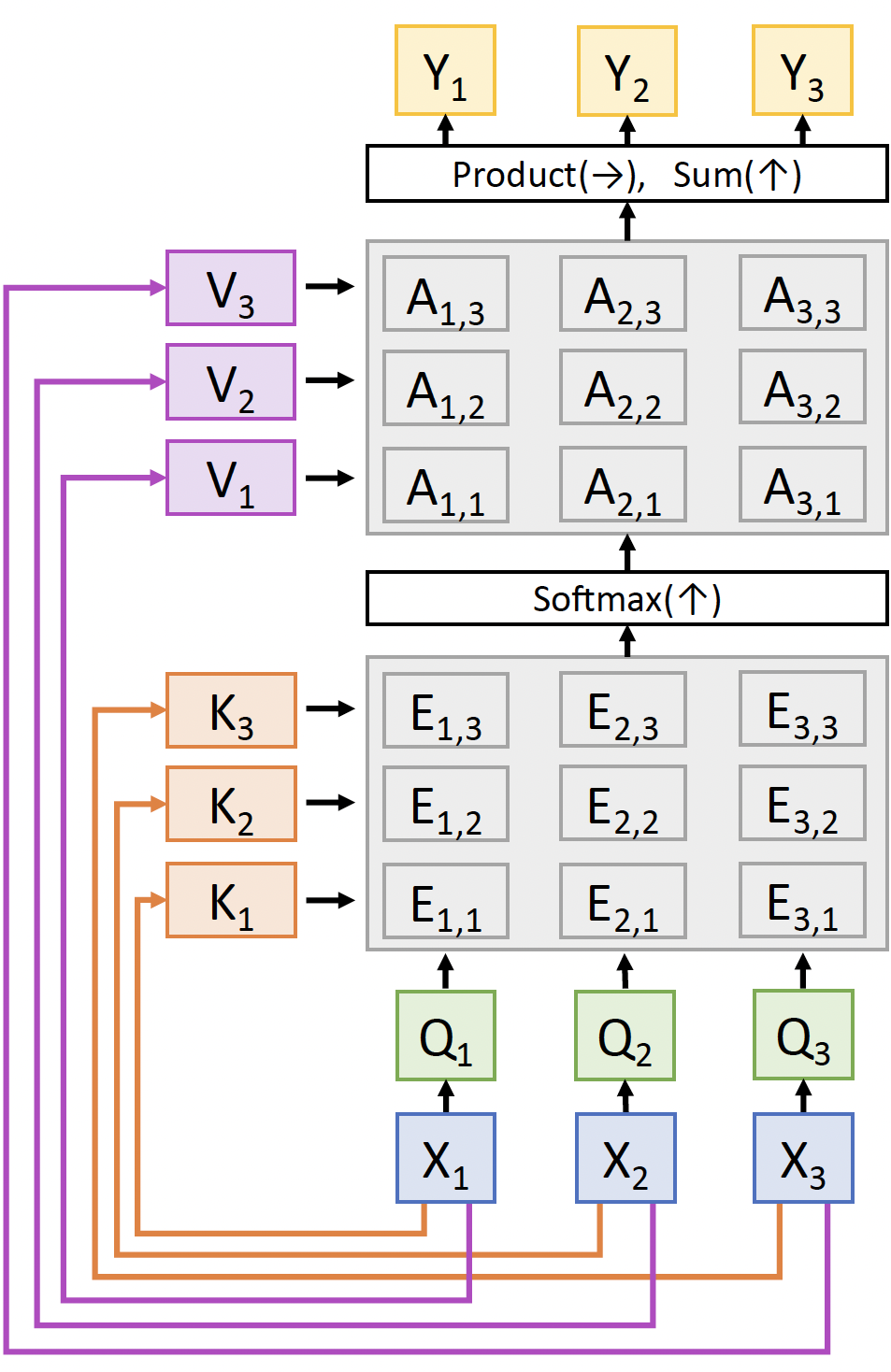}
    \vspace*{0.5mm}
    \caption{Overview of the self-attention module. Source:~\cite{transformer-umich}}
    \label{fig:self-attention-overview}
\end{figure}

In our work, we included both the original ViT and DeiT models, alongside the DeiT-Tiny to determine the effect of the network's size and complexity on its error resilience. For all transformers included in our studies, the architecture follows a pattern of stacked self-attention functions and fully connected layers. The self-attention functions can be thought of as mapping a query vector and a set of key-value pairs to a vector output, while the fully connected layers serve to project and modify the output vector to be used as input tokens to the next layer. Similar to the concept used in database retrieval, the vector output of self-attention is computed as a weighted sum of the input values, where the weight assigned to each value is computed by a compatibility function of the query with the corresponding key~\cite{vaswani2017}. As shown in Figure~\ref{fig:self-attention-overview}, the self-attention function first computes the corresponding query, key, and value matrices for the input tokens, and subsequently performs scaled dot-products between the generated matrices to arrive at the output vector. Since these GEMM operations make up the majority of the computation in vision transformers, our studies focus primarily on analyzing and protecting these operations. 


\noindent {\bf Formulation to Quantify Resilience.}
We focus on understanding the resilience of different transformers used in vision in the context of transient hardware errors at inference time. Specifically, we employ a single-bit flip error model that is commonly used as an abstraction for modeling hardware faults~\cite{ashraf2015understanding}. Within this error model, we focus on transient errors that successfully alter the originally correctly chosen classification class of an inference. We refer to this as a classification mismatch and it can be used as the baseline vulnerability metric for selective protection. 

CNN error resilience is a well-studied topic~\cite{hari2021making, ibrahim2020soft, er-inmemory-computing, Schorn2019automated-design, schorn2019bitflip, libano2019FPGA}.
Similar to a prior work~\cite{mahmoud2021optimizing}, we also formulated a transformer layer's vulnerability metric  based on the size of the layer and the likelihood of an error in the layer propagating to the output. Specifically, we define a layer's vulnerability as $V_{layer}[i] = V_{orig}[i] \times P_{prop}[i]$, where $V_{orig}[i]$ represents the probability that a transient hardware fault corrupts the output of the $i$-th layer and $P_{prop}$ represent the likelihood that the corrupted output will propagate to and modify the final model output. $V_{orig}[i]$ depends heavily on the hardware details such as raw failure rates, numerical precision, size of the layer (e.g., number of MACs and buffers used to execute the work in the layer). It can effectively be approximated as proportional to the number of MAC operations used to compute the output of the $i$-th layer, assuming that all MAC operations within the system have an equal likelihood of faulting and the large storage structures (e.g., FIFOs, buffers) are protected with ECC/parity. Without loss of generality, we compute and use the relative vulnerability for $V_{orig}[i]$ in this paper. It is computed as the number of MACs used in one forward pass of the $i$-th layer divided by the total number of MACs used in one forward pass of the entire model. By definition, $ 0.0 < V_{orig}[i] < 1.0$. 

A layer's propagation vulnerability, on the other hand, can be computed using either the number of mismatches that occur during our injection campaign of said layer, or a continuous metric we refer to as Delta Loss:
\begin{center}
$\Delta Loss_{layer} = \sum_{i=1}^{N}[\ell_{golden} - \ell_{i}]/N$ \\
\end{center}

where $\ell_{golden}$ is the cross-entropy loss for an error-free inference and $\ell_{i}$ stands for the cross-entropy loss for the corresponding $i$-th error-injected inference across $N$ total error injections. Similar to the case for $V_{orig}[i]$, the larger the value of $\Delta Loss_{layer}$, the more vulnerable the layer is to transient errors. The main advantage of $\Delta Loss_{layer}$ is that it converts the binary metric of classification mismatch into a continuous metric, which has been shown to significantly boost the speed of model analysis, especially when mismatches occur sparsely. 

\subsection{Related Work}
\label{seb:related_work}

For CNNs, different techniques have been explored to make inferences resilient. 
Exploding neuron values were identified as one of the primary reasons for inference output corruptions~\cite{Li2017, Chen2021}. A prior work profiled the value ranges in each layer during  fault-free executions and derived a range-based bounds checker to detect significant deviations~\cite{Li2017}. Ranger extended the concept and proposed an automated transformation to selectively restrict the value ranges in DNNs, which reduces the large deviations caused by critical faults and transforms them to benign faults that can be tolerated by the inherent resilience of the DNNs~\cite{Chen2021}. This work builds on top of these methods and only considers residual errors. We assume that such detectors or range restrictors are in place or the quantized data types used during deployment will inherently restrict the value ranges. We only consider errors that will perturb the value of the neuron within the profiled value ranges as computed using the training set. 

Mahmoud et al.\cite{mahmoud2021optimizing} introduced FILR, an overarching procedure that can be broken down into two separate resilience methods -- feature-map level resilience (FLR) and inference level resilience (ILR) -- as each method targets a different dimension when providing selective resilience for CNNs. During FLR, the network layers are ranked using a given vulnerability metric, and the layers are selectively protected such that it will simultaneously maximize the vulnerability coverage and minimize the overhead of redundancy. 
ILR, on the other hand, is used to determine the thresholds on network confidence to determine which inferences are vulnerable and need a rerun. 
Metrics to determine vulnerability for non-classification models (e.g., segmentation or detection) are not studied, which makes it difficult to apply ILR to such applications. While FLR is attractive (and has similarities to our solution), the duplication of layers as a protection method is too expensive and often results in a high-overhead solution. We overcome that challenge by leveraging a significantly lower-cost algorithm-based protection method.
\vasu{can we say above that our results show this?} \siva{Please edit the sentence above and I'll proof-read. We want to be somewhat suspenseful here and not reveal too much about work here as we want to first list out the questions and the reviewers should feel that - yes, we need answers for these.
In fact, I'm ok to remove all instances of "we solve this" before listing the questions. Feel free to edit as you think appropriate.}

Since most of the compute-intensive (convolution and fully connected) layers are linear operations and are executed as GEMMs (general matrix multiplications) on GPUs, algorithm-based fault tolerance (ABFT) techniques that verify and correct computation using checksums are applicable~\cite{huang1984algorithm}. These techniques compute checksums for input data, store them with the original data, perform the original and redundant computation, verify outputs, and correct errors inline. While the extra compute operations ABFT introduces are a small fraction of the number of the original computations, prior ABFT implementations have achieved about 20 percent runtime overheads for square matrices~\cite{Chong2011}. A recent study reported that the overheads can be much higher (${>}50$\%) for the non-square matrices that are typically used in CNNs~\cite{hari2021making}. 
Recent studies applied the approach to detect errors in convolutions~\cite{Filippas2022, hari2021making} and reported much lower overheads ($6{-}23$\%) without the inline error correction capabilities (referred as Algorithm-Based Error Detection or ABED). In complex safety-critical systems such as AVs, preventing silent data corruption (SDC) is more important than the ability to correct the error inline. Some of the prior solutions focus on fixed-point computations (e.g., int8 convolutions in~\cite{hari2021making}) where a precise checksum computation is viable and sufficient. In most practical applications, the use of floating-point data types (e.g., FP16/E5M10 or Bfloat16/E8M7) remains prevalent. Floating-point computations are known to introduce numerical error that makes checksum-based checking challenging. 

There is parallel work to ALBERTA that employs similar methodologies of providing transformer error resilience through checksum-based algorithms~\cite{10174239}. In our assessments, ALBERTA provides several advantages when compared to the alternative study. In terms of performance, ALBERTA achieves a substantial improvement in error recovery from 96\% to 99\%. When evaluating the experiment setup and results, ALBERTA evidently provides better versatility in handling production level architectures and dataset configurations. As opposed to analyzing CIFAR-10 with less relevant images in~\cite{10174239}, ALBERTA utilizes the full ImageNet dataset and allows for error injection campaigns of arbitrary sizes. Moreover, ALBERTA provides detailed studies of numerical discrepancies encountered during checksum computation, which is crucial to providing error resilience for floating pointing models. Combined with an easy extension proposed to address broader SDCs from hardware and systemic faults, the findings and error correction capabilities provided by ALBERTA translates directly to production settings characterized by substantial variations in inference data.


\subsection{Research Questions}
\label{sec:rq}
Given the challenges mentioned above, the primary objective of this paper is to answer the following four key research questions. 
\begin{itemize}
\item RQ1: How resilient are vision transformers to transient errors when performing image classification?
\item RQ2: What per-layer resilience patterns and variations exist, if any, for vision transformers? 
\item RQ3: Can we protect vision transformer models in a way to achieve high resilience with minimal computation and memory overhead?
\item RQ4: How do we design a reliable checksum-based error detection mechanism that maintains high error coverage for floating-point models? 

 
 \end{itemize}


\section{The ALBERTA Framework}
\label{sec:alberta}

We present ALBERTA, \textbf{al}gorithm-\textbf{b}ased \textbf{e}rror \textbf{r}esilience in \textbf{t}ransformer \textbf{a}rchitectures, a framework that allows us to analyze and improve the resilience of (vision) transformer models. We focus on vision transformers owing to their increasing use in safety-critical applications. We first present the vulnerability analysis of vision transformers in ALBERTA followed by our resilience technique.
 
\subsection{Vulnerability Analysis}

\begin{figure}[ht]
    \setlength{\belowcaptionskip}{-5pt}
    \setlength{\abovecaptionskip}{1mm}
    \centering
    \includegraphics[width=0.45\textwidth, clip]{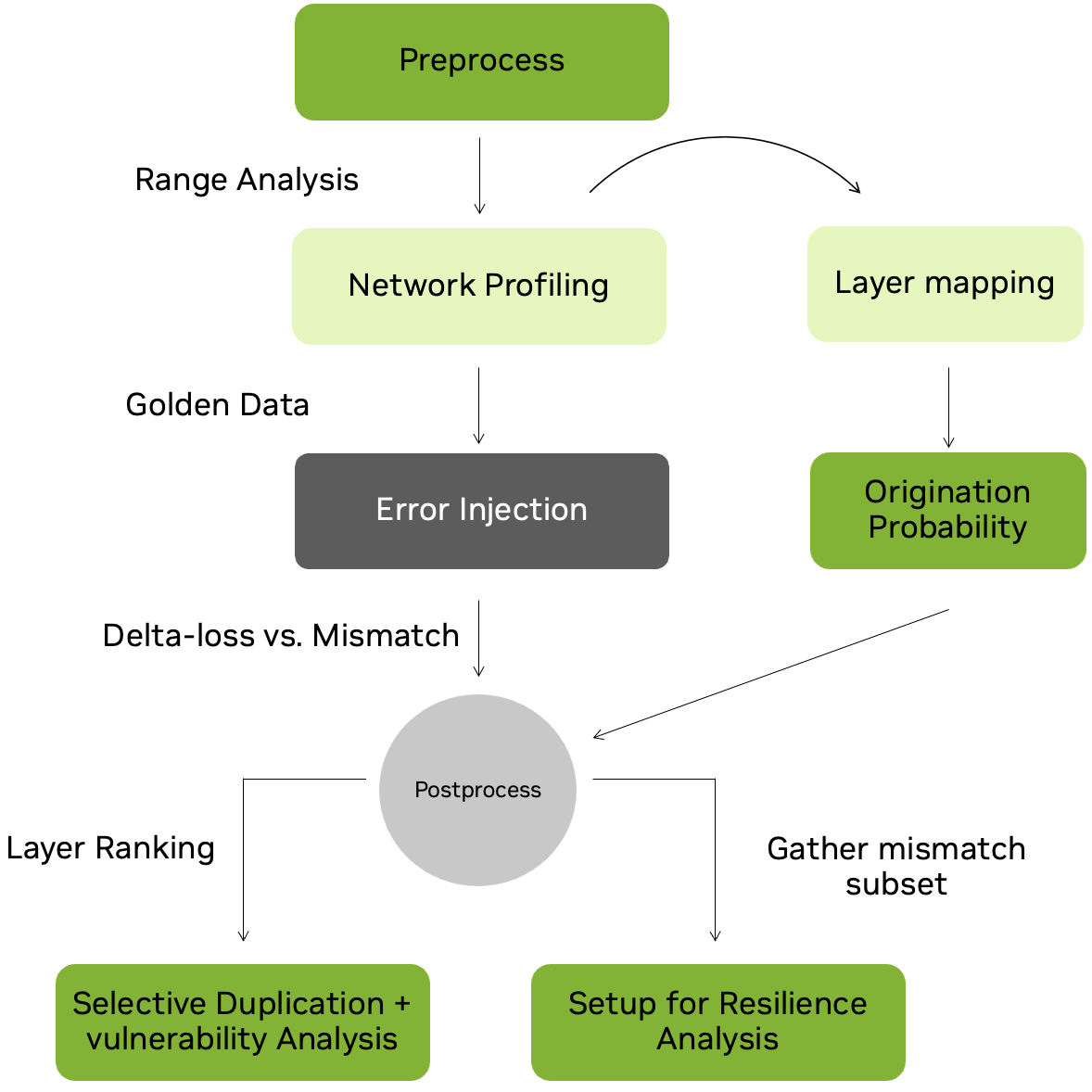}
    \vspace*{0.5mm}
    \caption{Pipeline Overview for ALBERTA vulnerability analysis}
    \vspace{-0.1in}
    \label{fig:pipeline_overview}
\end{figure}

 We quantify the vulnerability of individual layers with a pipeline that is divided into the following four stages. Figure~\ref{fig:pipeline_overview} summarizes the pipeline.

(1) The \emph{preprocess} step records the numerical lower and upper bounds for every layer in the network during inference to define the expected values ranges of neuron outputs. This information is later used in the error injection to rule out errors that can be protected by techniques like clamping and range restriction~\cite{Chen2021, Li2017}.
(2) The \emph{profiling} step records the network architecture details (layer sizes, parameter count, etc.) in addition to collecting the set of images that the error-free pre-trained model correctly classifies. We refer to this subset as \textit{golden data} and will use it as the sample space for error injection and resilience. 
 (3) The \emph{error injection} step performs data type dependent 
  random error injections for every layer within the network. Specifically, for every injection experiment, we flip a random bit at a randomly chosen neuron for a random image -- as constrained by the numerical range determined during the preprocess step -- and record the location of the error injection with the corrupted output. 
(4) The \emph{postprocess} step compiles the results from injection and profiling steps into pandas dataframes for selective layer level duplication.

In the error injection step, we perform 102400 injections for very low error in the measurement at high confidence. With the above sample size, the error bar is ${<}0.24$\% at $99$\% confidence level (using $90$\% as population proportion). Depending on the desired error tolerance in the results, fewer injections may be sufficient. During each individual error injection, we sample from the set of Golden Data containing 5017600 images that the model correctly classifies and select batches of size 128 or user specified number of images for analysis. For each selected image, the output activation of the injected layer is modified using the target injection type and the corrupted classification loss -- alongside other injection metadata -- is recorded during inference for downstream analysis.  

With the objective of protecting the model with the lowest possible overhead, we focus on using the vulnerability data collected from the pipeline to rank the network layers, such that employing a chosen protection algorithm, such as selective layer duplication or checksum style error detection, will simultaneously maximize the resilience and minimize the overhead of redundancy. In the case for selective layer duplication, the coverage provided by protecting a target layer is computed using the equation for $V_{layer}$, while the overhead of duplicating a target layer depends on the amount of work in the layer (e.g., number of MACs, which is proportional to $V_{orig}$ according to the formulation described in Section~\ref{sec:background}).

The majority of the computation in transformer architectures is performed in linear (fully connected) 
layers that are a part of each transformer block. To provide comprehensive error protection, we select four layers to be included in our analysis pipeline for every transformer block. This consists of the linear layer used to compute key query value matrices (Q, K, V) as shown in Figure~\ref{fig:self-attention-overview}, the linear projection layer that outputs the result of multi-head attention, and two accompanying linear layers used by the feed-forward MLP. The matrix multiplication that occurs within the self-attention computation (dot product between query and key matrix and subsequently the product between the softmax output and the value matrix) are not considered in this approach and will be addressed in future work due to user-end implementation complexity.
As of Pytorch version v2.0.0, all efficient error injection methods are done through Pytorch hooks, whose applications are limited to well defined Pytorch modules as opposed to general operations such as matrix multiplication. Although it is possible to use a user-defined self-attention module with custom matrix multiplication layers, doing so would render the framework not generalizable where each target model must be individually modified.

\subsection{Resilience Analysis}\label{sec:experiments}


This section presents our resilience improvement technique in the layers identified as vulnerable based on the above analysis. We start with the description of the Pytorch wrapper class that
integrates ALBERTA into different transformer models. Next, we describe the execution and evaluation process which includes layer selection, parameter tuning, and generating the false positive and negative rates for error detection. Lastly, we provide the details of the correction mechanism that can be deployed alongside ALBERTA 
for forward error correction to complete the end-to-end protection solution.

\subsubsection{Resilience Implementation}
ALBERTA's resilience implementation is defined as a Pytorch module. It provides a set of functionalities that enables an in-depth resilience analysis of transformer based architectures. Apart from the basic utility functions such as custom loading of models and datasets, the resilience implementation contains several key functionalities that sets it apart from previous works as listed below. \siva{It felt like a numbered list should immediately follow this sentence.}

\begin{figure}[t]
    \setlength{\belowcaptionskip}{-5pt}
    \setlength{\abovecaptionskip}{1mm}
    \centering
    \includegraphics[width=0.4\textwidth, clip]{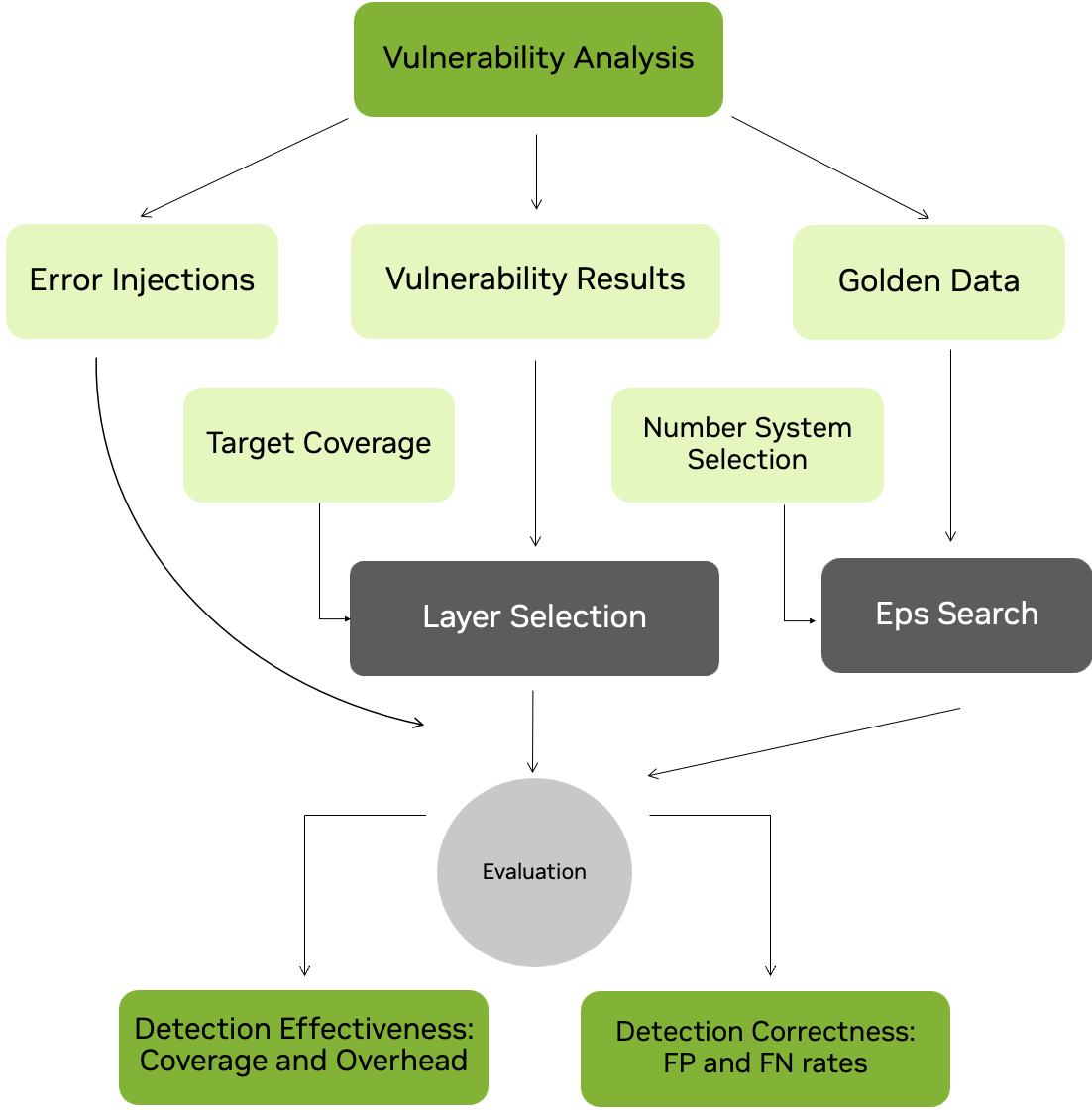}
    \vspace*{0.5mm}
    \caption{Pipeline Overview for ALBERTA's resilience implementation}
    \label{fig:abed pipeline overview}
\end{figure}

\begin{figure}
    \setlength{\belowcaptionskip}{-5pt}
    \setlength{\abovecaptionskip}{1mm}
    \centering
    \includegraphics[width=0.5\textwidth]{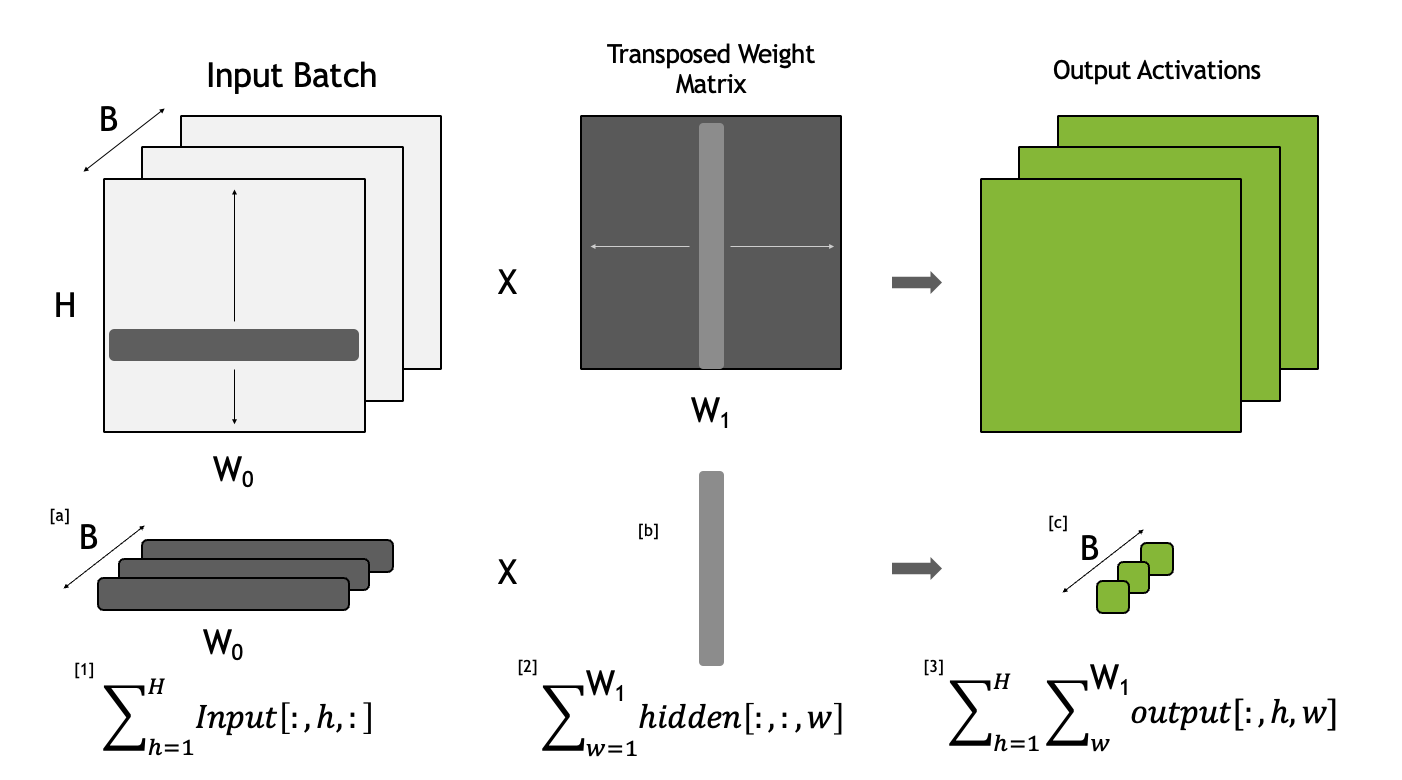}
    \caption{The error detection checksum compares the value of step-3 against the green vector in step-c element-wise for any numerical discrepancy and computes the batch average difference for distribution modeling during eps search. }
    \vspace{-0.1in}
    \label{fig:abed checksum overview}
\end{figure}

\noindent\textit{Checksums.} The core function of ALBERTA is to enable checksum style protection for the subset of layers that are deemed most vulnerable according to prior layer level analysis. In the case for most transformer models, such subsets consists solely of GEMM layers whose output activations are obtained by performing batched matrix multiplication between the input features and the transposed hidden weight matrix, while also adjusting the result by the layer bias. As part of ALBERTA's core functionality, we perform checksum style verification on this general matrix multiply (GEMM) operation and successfully avoid the overhead of traditional ABFT checksum matrices. To accomplish this, we forfeit the capabilities of inline error correction and instead compute row and column checksum vectors for the input and layer activation, respectively. Specifically, as depicted in Figure~\ref{fig:abed checksum overview}, the batched row checksum vector is computed by summing along the column dimension of the input batch, while the column checksum vector is computed by reducing the row dimension of the transposed hidden weight matrix. To perform error detection, we compare the dot product of the row and column checksums against the sum of layer activations for numerical discrepancy. It is worth noting that (1) this process offers a significant overhead reduction when compared to traditional techniques such as fmap duplication and modular redundancy -- as shown later in the results section and (2) it can be optimized such that the column checksum of the transposed hidden weight matrix can be computed offline and stored by ALBERTA's resilience implementation prior to deployment.

While the checksum itself can be naively implemented using torch tensor operations, some challenges must be addressed for this detection mechanism to be generalizable.

\noindent (1) As introduced in the findings of~\cite{hari2021making}, the increasing use of reduced-precision data types (e.g., 8- and 4-bit integers) during inference accelerators introduces new challenges for checksum-based error detection techniques, as successive sums across large layer activations have a high likelihood for overflow. Although floating-point data types do not exhibit the extreme overflow behaviors observed in integers, significant numerical imprecision still occurs in the case where depending on the rounding mode -- overly large results will be represented as max float (RTZ) or Inf (RNE), while underflown results are often rounded to zero. In situations where such rounding errors necessitate significant increases in the discrepancy tolerance levels used during error detection, ALBERTA's resilience implementation includes additional functionalities for automatically selecting the appropriate numerical precision used for checksum computations based on the numerical range of network weights and the input dataset.

\noindent (2) An equality check is sufficient when dealing with models that use fixed point data types. However, a more nuanced checksum verification is needed for floating-point models, as discrepancies in the checksum can arise due to the nondeterministic behavior of floating-point arithmetic, as opposed to transient errors. ALBERTA's resilience implementation addresses this issue by providing an intuitive, confidence based mechanism for identifying whether a discrepancy is caused by transient error or numerical imprecision. We now describe this epsilon generation process.

\noindent\textit{Eps Generation.} One significant limitation of prior works in algorithm based fault tolerance (ABFT) is the restricted application to fixed datatypes where the checksum computations are deterministic. Our work provides a solution to this issue by introducing a confidence based error detection framework capable of achieving close to 100\% coverage of transient errors that result in model misclassification. In order to accomplish this, ALBERTA's resilience implementation includes a numerical analysis function that empirically models – for each layer –  a unique distribution describing the probability that a discrepancy in the layer’s online checksum is due to numerical imprecision rather than transient error. More specifically, this process involves first iterating through the subset of data that the error free network correctly classifies, while collecting during every inference a set of differences between the computed checksum and the true layer activation sum. Using this set of discrepancies which are unique to each layer’s trained parameters and dimensions, a layer specific distribution can then be generated through regression and used for error detection during deployment, where if the observed checksum discrepancy is above a user set confidence threshold, an error would be raised and the correction mechanism would be triggered.

\subsubsection{Execution and Evaluation} Aside from allowing the users to select their desired numerical precision for the checksum independently from the model itself, ALBERTA offers two additional tuning mechanisms that the user can use prior to the evaluation step:
\begin{enumerate}
\item The option to either select custom layers for resilience or provide a desired theoretical error coverage and have ALBERTA's resilience implementation dynamically select the layers necessary for achieving it.
\item Selecting the target distribution (defaults to standard normal) and confidence threshold of checksum discrepancies based on which error detection will be performed. 
\end{enumerate}

As we show in the vulnerability analysis, it is not uncommon for transformer architectures to have layers that are extremely resilient to transient errors. As a result, we can reduce overhead by omitting these layers. By simply providing a desired coverage level, ALBERTA's resilience implementation will automatically perform this optimization through a knapsack inspired mechanism. It is important to note that since ALBERTA behaves independently for each layer of the network, selecting a subset of layers for protection means error occurring in any unprotected layers cannot be detected; making it important to select layers with the highest error propagation rate and overall vulnerability for protection. In other words, when we decide to not include a layer for checksum protection, we are choosing to forefeit all error coverage for that layer in exchange for a reduction in network overhead. Furthermore, the optimal subset of layers returned by ALBERTA is based on estimated layer vulnerabilities that are dataset dependent. As a result, the resilience bounds we provide in the results section are not strictly true when fine-tuning models on new target datasets, but can always serve as a statistically accurate approximation.

\subsubsection{Correction} 
In addition to providing the necessary tools to integrate and evaluate the effectiveness of checksums as a protection mechanism for arbitrary transformer models, the wrapper also offers the choice of two correction mechanisms to be performed after error detection – replay and skip forward.

\noindent To perform a \emph{replay} upon error detection, we first save the activation values of every layer during inference, and if the subsequent layer raises an error, we simply read the error-free activations from the previous layer and rerun the inference step. This mechanism incurs constant space overhead -- since only one activation needs to be saved at a time during evaluation -- while also resulting in minimal runtime overhead as shown later in the results section. 

\noindent To perform a \emph{skip forward} upon error detection, we have the following three options – jump ahead to the next layer with the same input size, skip to the next transformer block, or simply jump to the inference step while passing forward the saved, error free activation as input. This mechanism can be used for any layer inside the transformer blocks with zero overhead but will likely have a negative impact on the model accuracy. For this reason, we believe it is applicable only for special use cases and have designed ALBERTA's resilience implementation to default to replay.

\section{Results}
\label{sec:results}
\subsection{Evaluation Methodology}
We perform our evaluation with a focus on DeiT-base and DeiT-tiny models, which are both pretrained and fine-tuned with distillation on ImageNet-1k (1 million images, 1,000 classes). We use the Pytorch Framework and obtained pretrained models from the Facebook research repository with patch 16 and 224x224 pre-training resolution.  All experiments are run on a standard Linux server with NVIDIA A6000 GPUs and Intel i9-10980XE CPU. We obtained the results for vulnerability analysis using an adapted version of the GoldenEye injection module \cite{mahmoud2022goldeneye} and perform all other testing using custom Pytorch hooks. During vulnerability analysis, all inferences are performed using full fp32 precision, while inferences during resilience analysis are executed with half-precision at fp16. We evaluate checksum operations comprehensively at fp16, fp32, and fp64. However, due to limitations mentioned regarding checksums, the best accuracy during error detection are obtained using fp64 and are included here in both overhead and coverage related plots.

\subsection{Vulnerability Analysis}
\label{results:analysis}

We first present the results from ALBERTA's vulnerability analysis on DEiT models.
We excluded the original ViT results as they are similar to base DeiT. We note that the original ViT performs significantly worse than its DeiT counterpart -- its error propagation rate and (as a result) vulnerability exceeds that of DeiT across all layers in the network. The cause for this behavior can be attributed to the network parameters learned during distillation versus regular training. 

\begin{figure}
    \centering
    \begin{subfigure}[b]{0.4\textwidth}
        \centering
        \text{DeiT-base}\par\medskip
        \vspace*{-0.2cm}
        {\includegraphics[width=\textwidth]{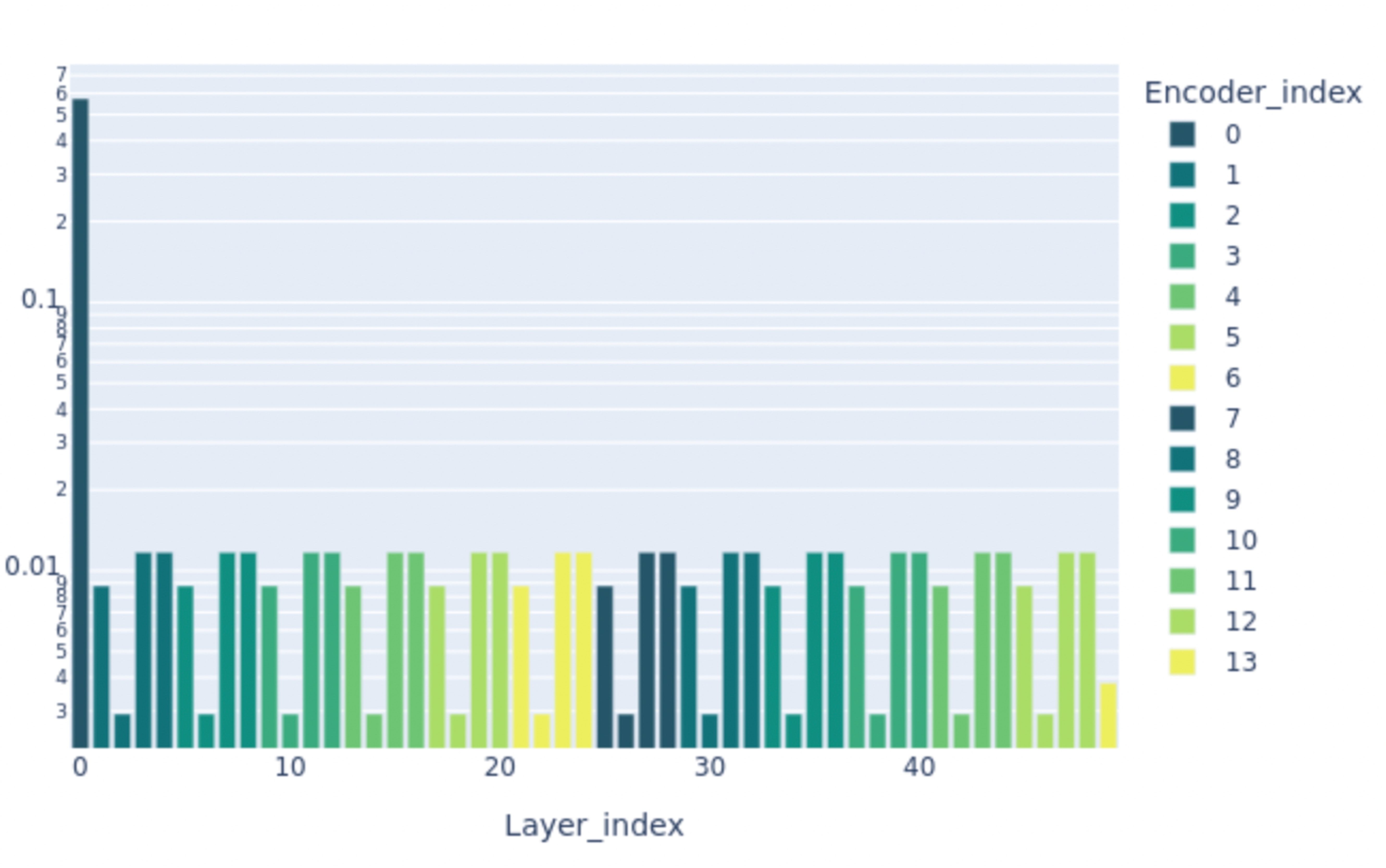}}
    \end{subfigure}
    \vspace*{-0.28cm}
    \begin{subfigure}[b]{0.4\textwidth}  
        \centering 
        \text{DeiT-tiny}\par\medskip
        \vspace*{-0.2cm}
        {\includegraphics[width=\textwidth]{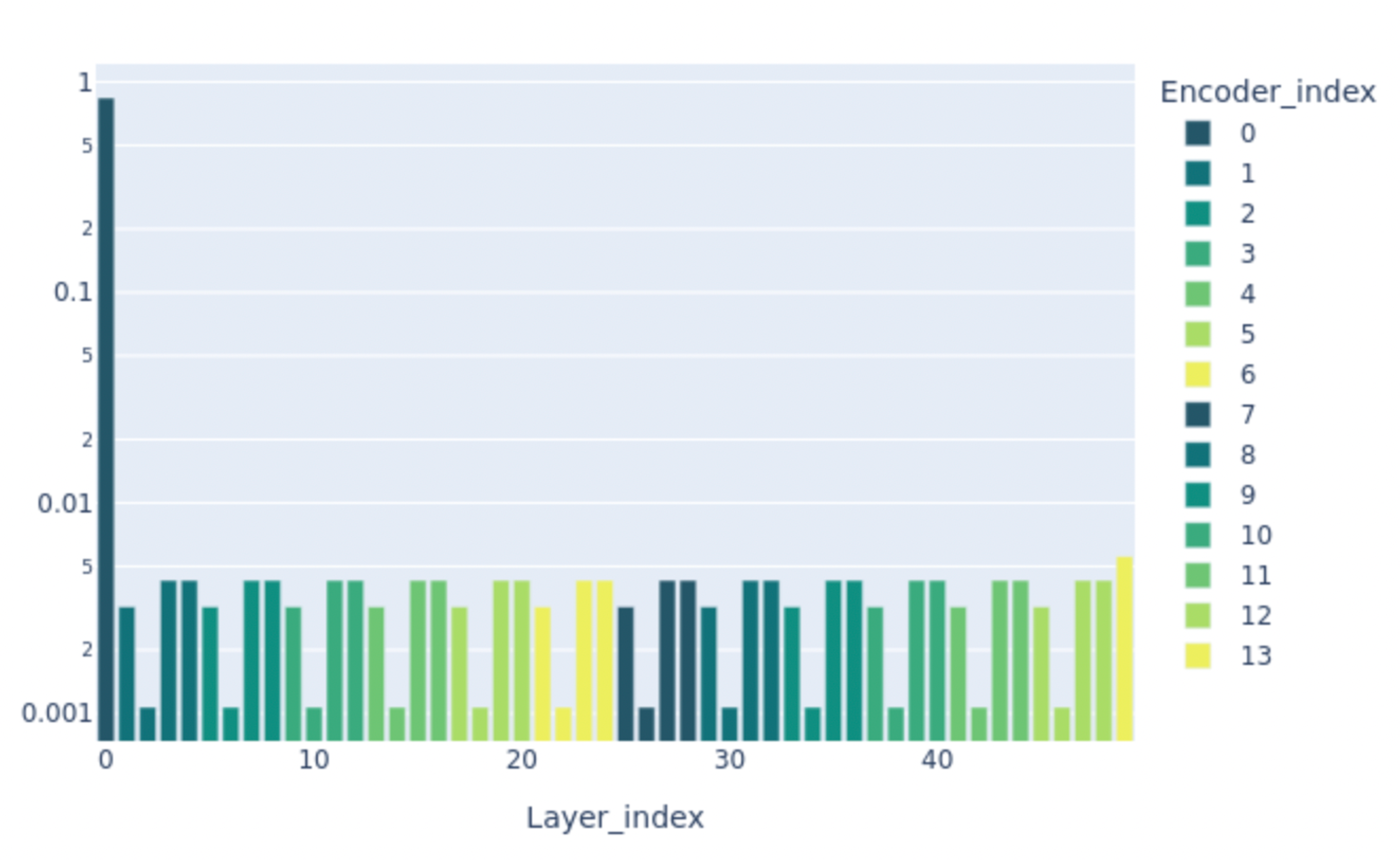}}  
    \end{subfigure}
    \caption[V-orig Values]
    {$V_{orig}[i]$ computed as the fraction of MAC operations used in the $i$-th layer over the total number used by the network. 
    The x-axis shows the layer number and the y-axis shows the $V_{orig}$ value.
    }
    \vspace{-0.2in}
    \label{fig:metrics on original data}
\end{figure}

\vspace{0.05in}
\noindent\textbf{Origination Probability Comparison.}
During the profiling step of our pipeline, we compute $V_{orig}$ as the fraction of MAC operations used in the forward pass of each layer (shown in Figure~\ref{fig:metrics on original data}). We observe that the largest value corresponds to the positional embedding layer (convolution), which is about 100$\times$ bigger than the intermediate layers inside the transformer block. Furthermore, the layer sizes are proportionally identical among both models included in our study. 

\begin{figure*}[hb]
    \centering
    \captionsetup[subfigure]{justification=raggedright}
    \captionsetup[subfigure]{justification=raggedright}
    \begin{subfigure}[b]{0.32\textwidth}  
        \centering 
        {\includegraphics[width=\textwidth]{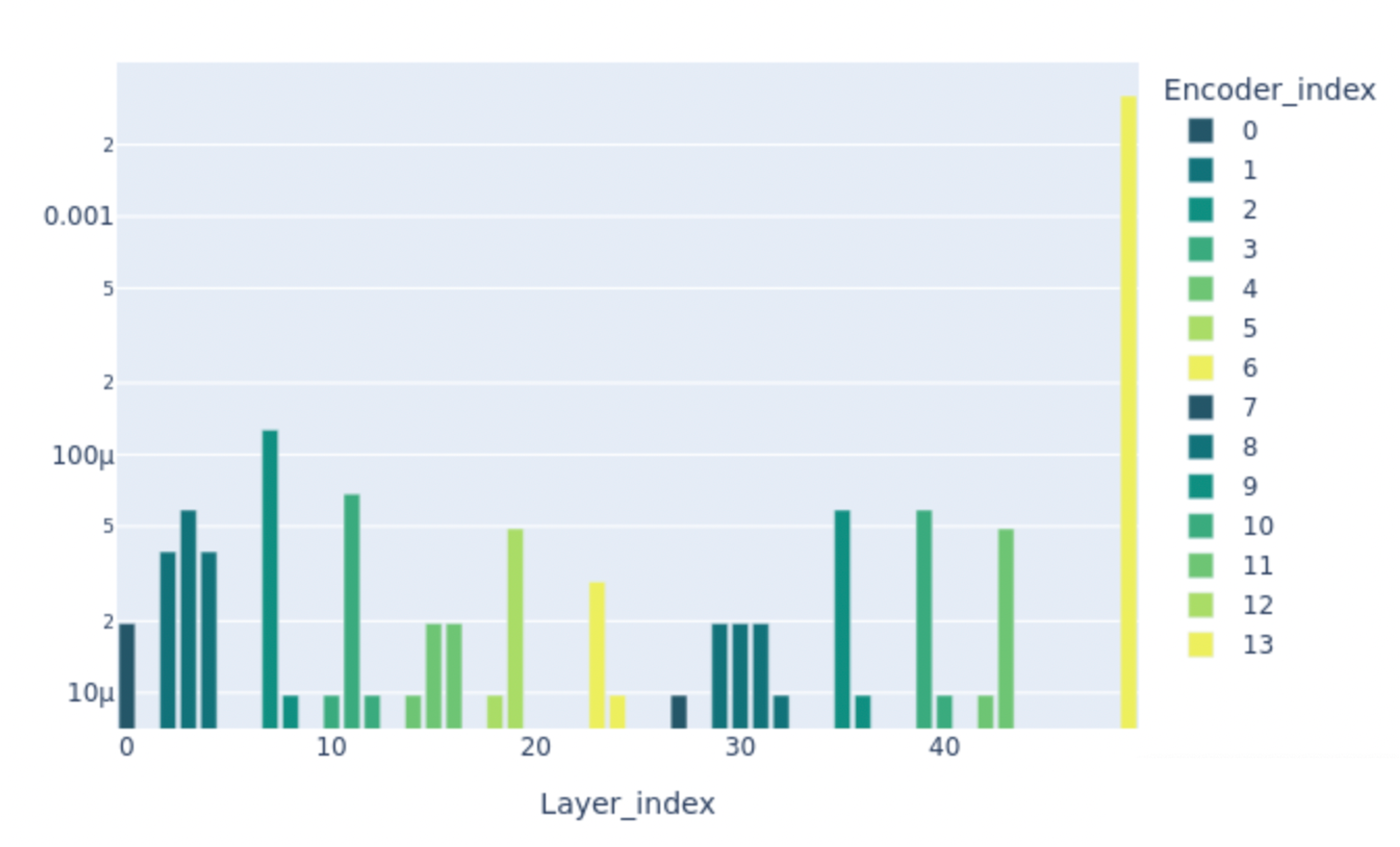}}
        \caption{DeiT-base}
    \end{subfigure}
    \captionsetup[subfigure]{justification=raggedright}
    \begin{subfigure}[b]{0.32\textwidth}  
        \centering 
        {\includegraphics[width=\textwidth]{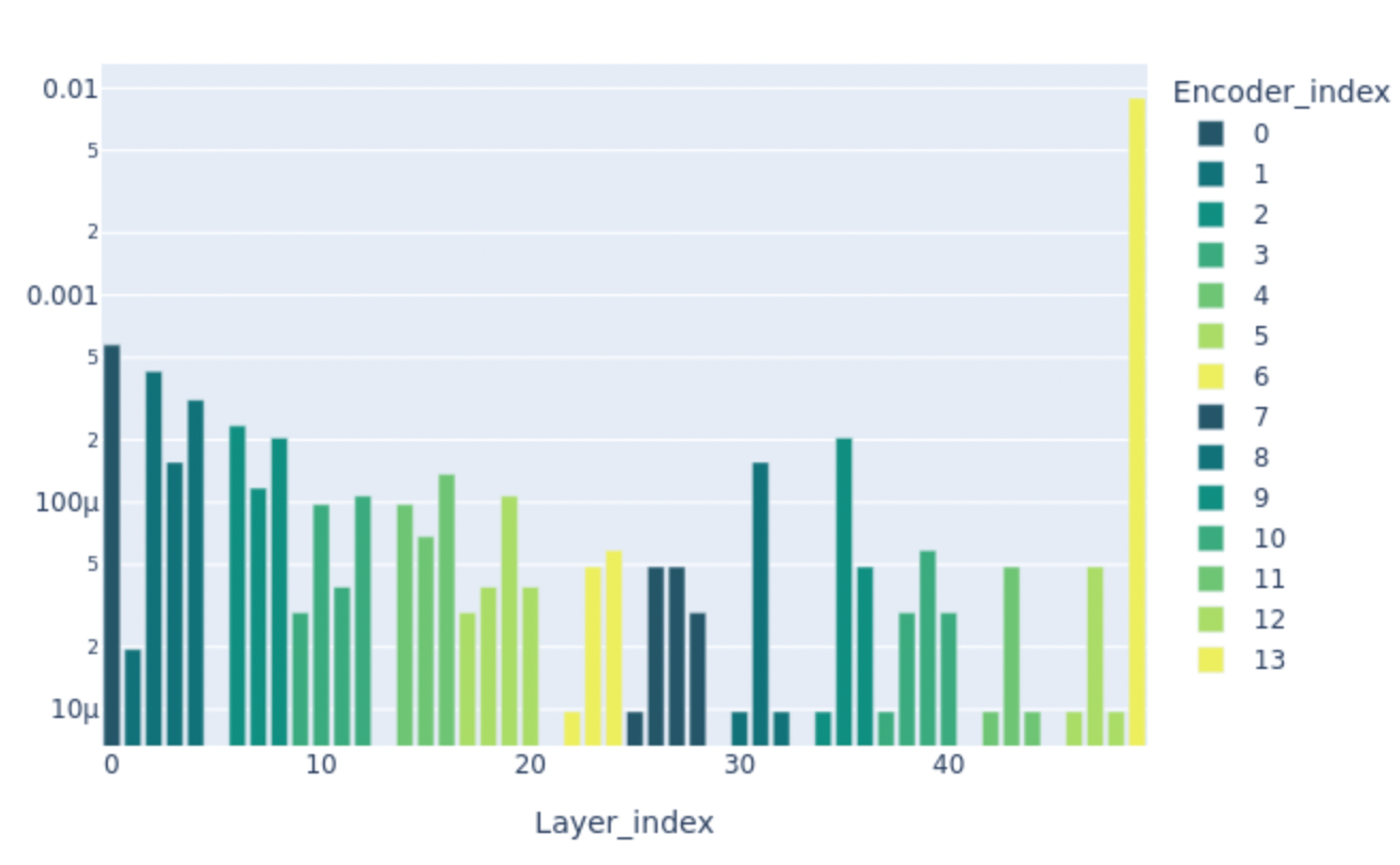}}
        \caption{DeiT-tiny}
    \end{subfigure}
    \begin{subfigure}[b]{0.32\textwidth}  
        \centering 
        {\includegraphics[width=\textwidth]{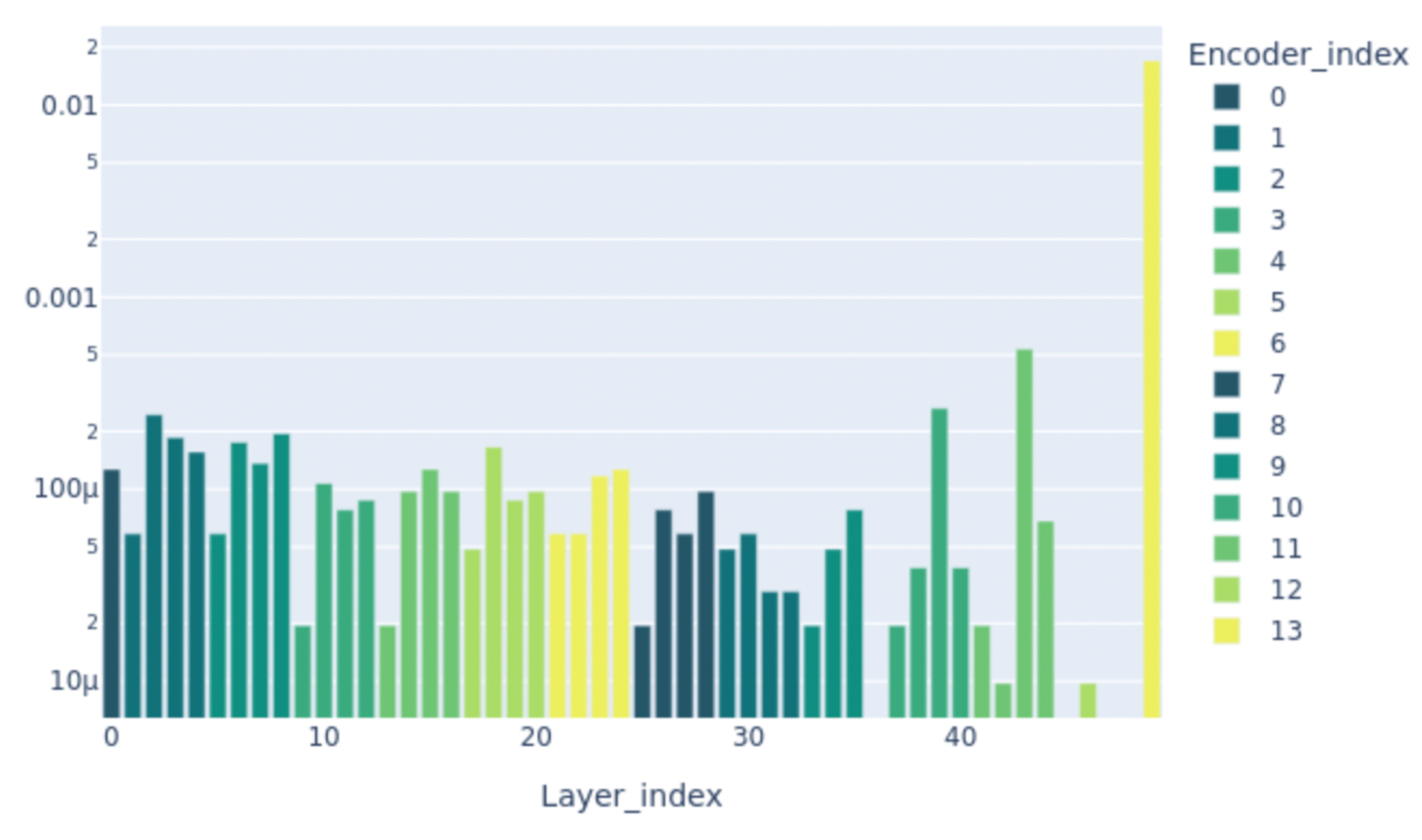}}
        \caption{ViT-Base}
    \end{subfigure}
    \caption[V-prop-mismatch]
        {$V_{prop}[i]$ as computed using mismatch. The x-axis represents the layer number within the corresponding architecture, while the y-axis represents the $V_{prop}$ value}
    \vspace{-0.2in}
    \label{fig:v-prop}
\end{figure*}
\vspace{0.05in}
\noindent\textbf{Propagation Probability Comparison.}
Using mismatch as the ground truth vulnerability metric for computing $V_{prop}$, we note that both the DeiT models exhibit a similar behavior where error injections performed at the final linear prediction head resulted in the highest probability of mismatch in the output. This is illustrated in Figure~\ref{fig:v-prop}.

More importantly, it can be observed that a significant number of layers within the DeiT-base model are incredibly resistant to transient errors. Moreover, DeiT-base has significantly better built-in resilience to transient error propagation when compared to DeiT-tiny. Out of 102400 error injections, zero random bit flip resulted in mis-classification of the network output for 22 out of the 50 layers. Comparisons between the results of DeiT-base and DeiT-tiny support the conclusion that a larger network size contributes to a lower error propagation probability. In practice, this behavior can be intuitively understood as the result of larger transformer networks containing higher number of normalization and activation layers that can likely suppress errors during inference.

\begin{figure}
    \centering
    \captionsetup[subfigure]{justification=raggedright}
    \begin{subfigure}[b]{0.4\textwidth}  
        {\includegraphics[width=\textwidth]{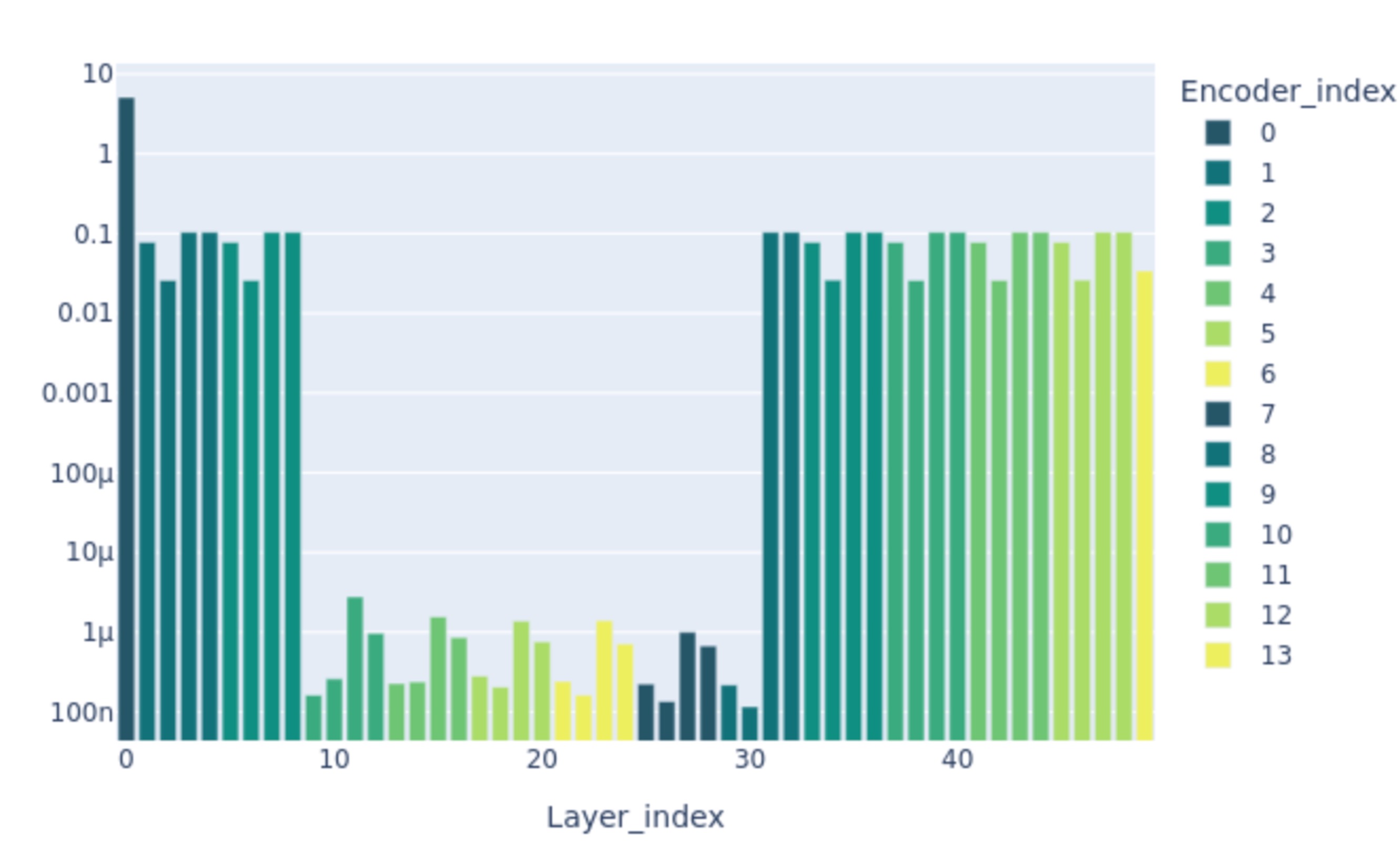}}
        \vspace{-0.2in}
    \end{subfigure}
    \caption[$V_{layer}$]
        {$V_{layer}$ of DeiT-base computed as $V_{orig}[i] \times P_{prop}[i]$}
    \label{fig:vlayer-deit}
\end{figure}

\vspace{0.05in}
\noindent\textbf{Delta Loss.}
$\Delta Loss$ converts binary classification mismatch into a continuous metric. We observe that the error propagation graph of DeiT-base exhibits a considerable dip in $V_{layer}$ value starting from the 3rd to the middle of the 8th transformer block as shown in Figure~\ref{fig:vlayer-deit}. Although this is consistent with the claim that DeiT-base has significantly better resilience than its DeiT-tiny counterpart, we do not fully understand how the random sparsity difference in the Mismatch graph is transformed into a contiguous block of minimal vulnerability in the DeiT-base architecture. Keeping in mind that both delta loss and mismatch measure the same latent layer vulnerability, we propose alternate error models based on finer-grained error injections as part of future investigation into this statistical divergence.

\begin{figure}
    \centering
    \captionsetup[subfigure]{justification=raggedright}
    \begin{subfigure}[b]{0.45\textwidth}  
        {\includegraphics[width=\textwidth]{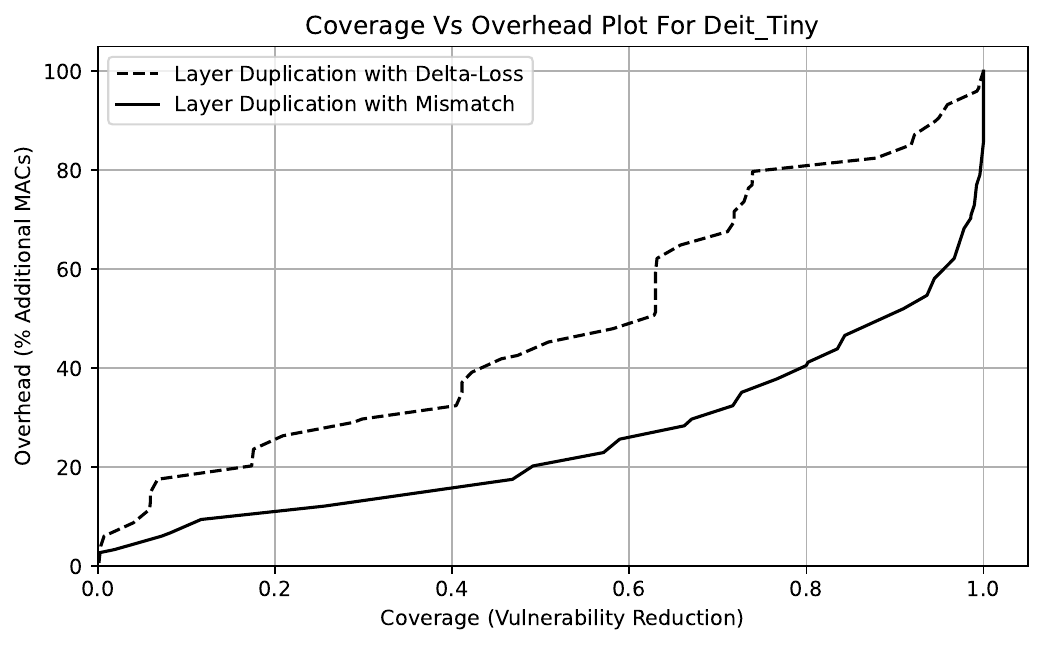}}
    \end{subfigure}
    \caption[Coverage Overhead graphs]
        {The coverage result of DeiT-tiny as computed from optimized selective layer duplication; maximizes the vulnerability / coverage of selected layers while minimizing the overhead of redundancy. The prediction head is not considered during this evaluation, as it should always be protected}
        \vspace{-0.1in}
    \label{fig:DeiT-coverage-overhead}
\end{figure}

\vspace{0.05in}
\noindent\textbf{Selective Layer Duplication}
%
%
%
The coverage versus overhead results from selective layer duplication for DeiT-Tiny are shown in Figure~\ref{fig:DeiT-coverage-overhead}, where x and y axes represent the total coverage and cumulative overhead, respectively. The dotted line represents the result of using the $\Delta Loss$ derived layer vulnerability, while the solid oracle line is the result of using mismatch. Due to the prediction head of transformer architectures being consistently the most vulnerable layer (as measured using mismatch), we assume it is always protected by the algorithm and instead plot the coverage vs. overhead from protecting the remaining layers. 

We present here two main takeaways from this study using Mismatch as the ground truth vulnerability. First, using the intersection between x = 90\% and the coverage graphs, it can be observed that DeiT-Tiny is able to achieve 90\% of the remaining vulnerability coverage at approximately 49\% overhead. Second, a significantly steeper overhead can be seen when trying to obtain the remaining 10\% coverage. To reach 99\% coverage for instance, select layer duplication would incur approximately 45\% additional overhead, which is prohibitively high in most high performance systems. While 90\% may be sufficient for architectures that have high fault tolerance, safety critical applications often demand higher reliability that are closer to full 100\% coverage. As a solution to this shortcoming, we introduce the alternative checksum-based protection for vision transformers. 

\subsection{Protecting Vision Transformers.}
We now present the main results from our experiments on error resilience, as divided into the following points of focus.  
(1) The empirical statistics of numerical discrepancies observed during floating-point checksums.
(2) The coverage and overhead from selecting optimal subsets of layers for checksum protection.
(3) The overall effectiveness of ALBERTA's confidence-based error detection and correction.

\begin{figure}[t]
    \centering
    \begin{subfigure}{0.24\textwidth}
        \includegraphics[width=\textwidth, clip]
        {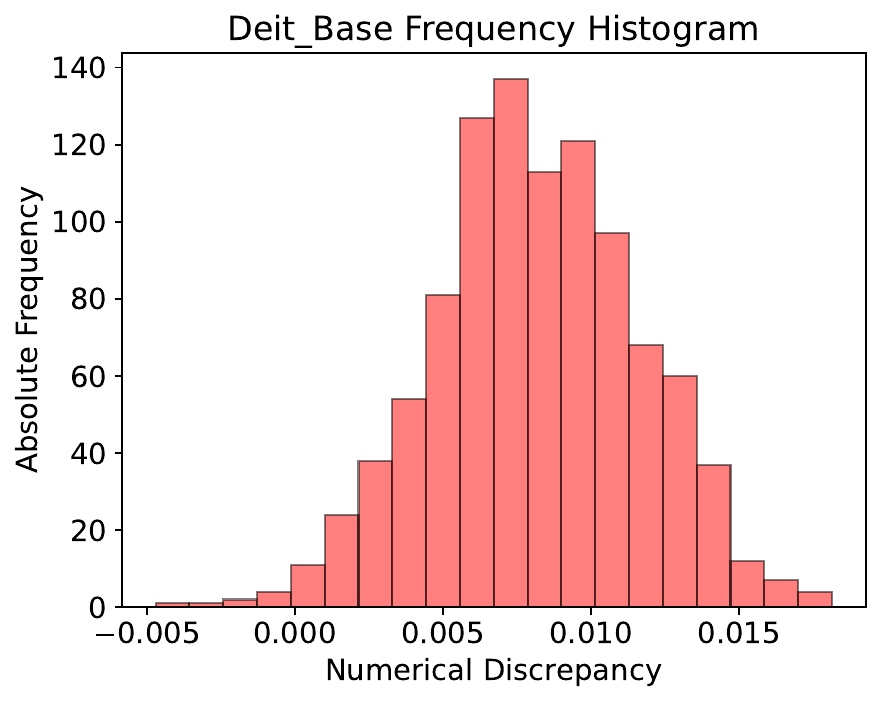}
        \caption{DeiT-base}
    \end{subfigure}
    \begin{subfigure}{0.24\textwidth}
        \includegraphics[width=\textwidth, clip]
        {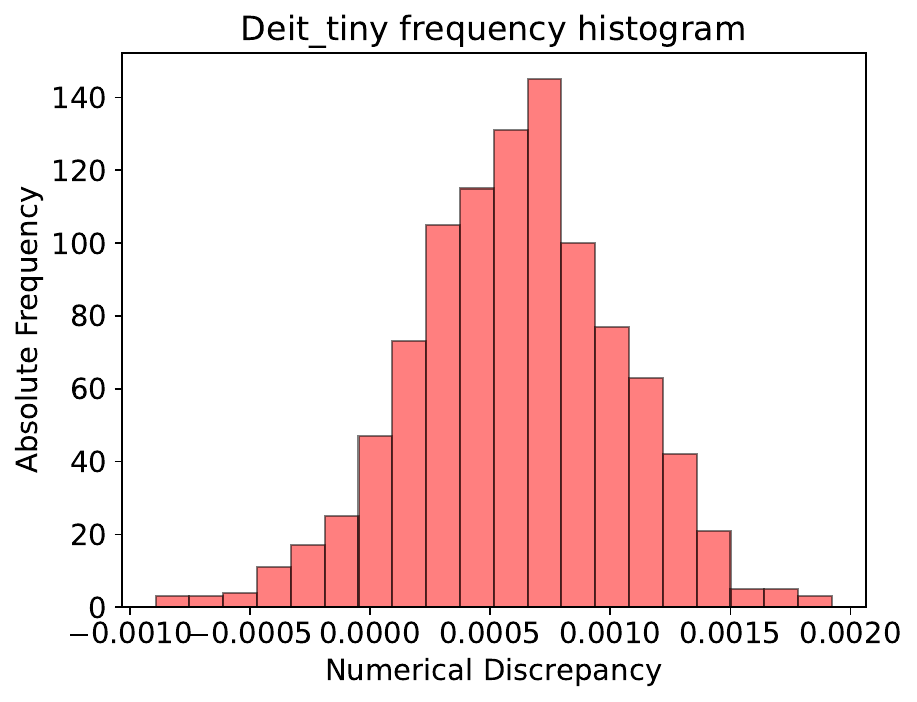}
        \caption{DeiT-tiny}
    \end{subfigure}
    \setlength{\belowcaptionskip}{-5pt}
    \vspace{-0.2in}
    \caption{Histogram of sample layer's discrepancy values in DeiT-base and DeiT-tiny}
    \label{fig:abed layer eps}
\end{figure}

\vspace{0.05in}
\noindent\textbf{Empirical statistics of numerical discrepancies.}
At its core, the basis for using confidence-based error detection is the observation that the empirical discrepancies encountered during checksum form a normal-like distribution for all Imagenet samples at inference. As shown in Figure \ref{fig:abed layer eps}, the same linear layer in DeiT-base and DeiT-tiny -- sizes (768,3072) and (192, 768) -- both exhibit a normal-like behavior, with the overall size decrease of 16x (3072 / 192) translating to a similar magnitude drop in observed discrepancies.

\begin{figure}[t]
    \centering
    \begin{subfigure}{\textwidth}
        \includegraphics[width=0.4\textwidth, clip]
        {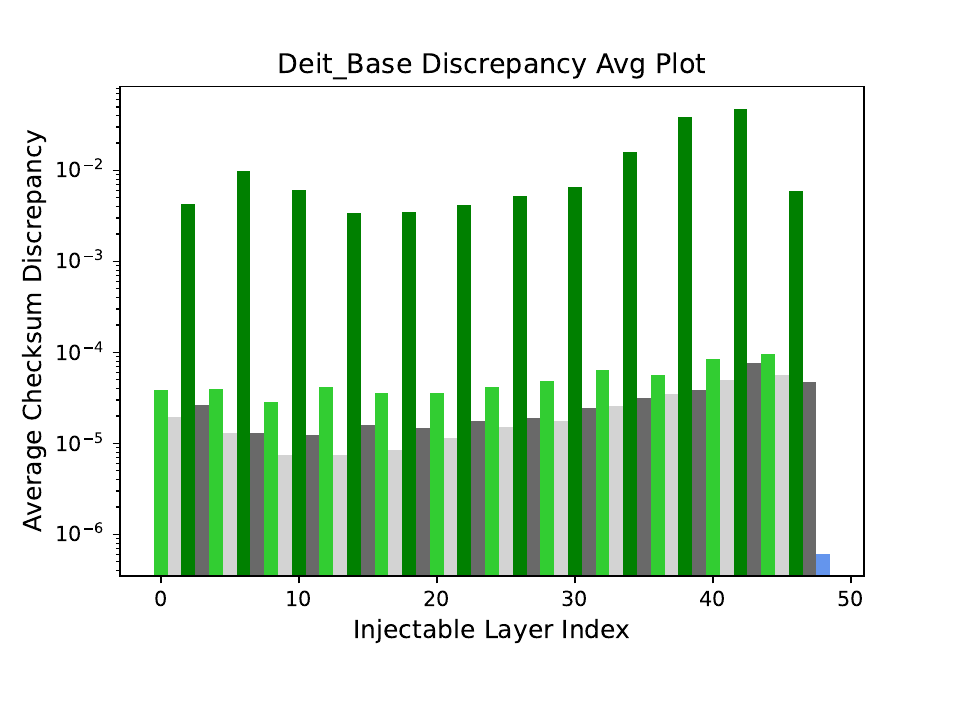}
    \end{subfigure}
    \begin{subfigure}{\textwidth}
        \vspace{-0.2in}
        \includegraphics[width=0.4\textwidth, clip]
        {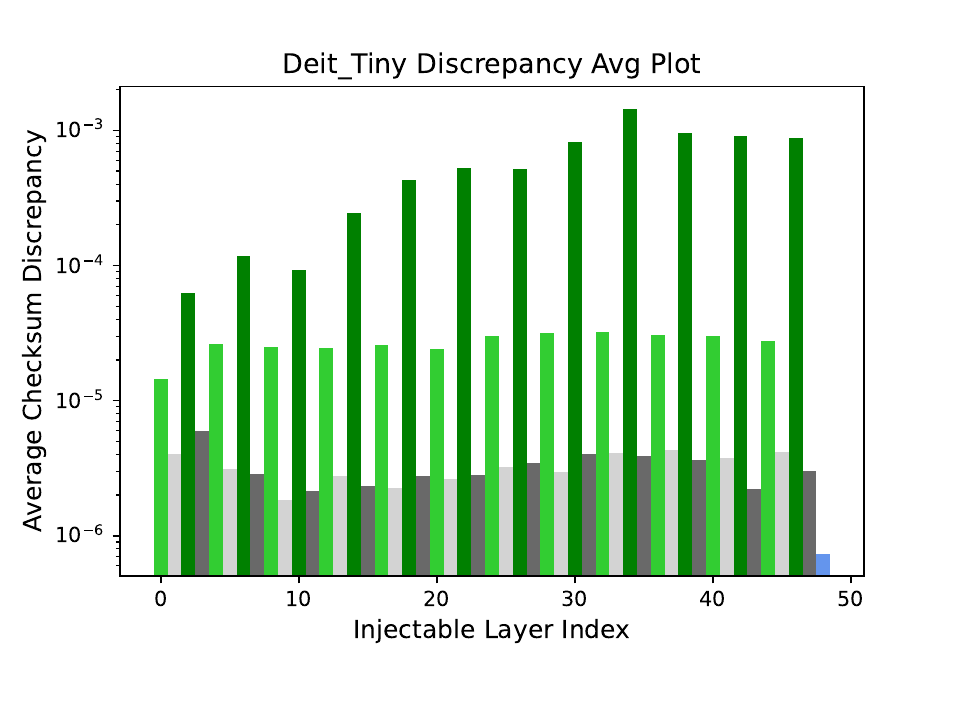}
    \end{subfigure}
    \setlength{\belowcaptionskip}{-5pt}
    \setlength{\abovecaptionskip}{-1mm}
    \vspace{-0.25in}
    \caption{Average checksum discrepancies observed during Epsilon Search for DeiT-base \& DeiT-tiny}
    \vspace{-0.1in}
    \label{fig:network avg eps}
\end{figure}

At the network level, similar trends can be seen in figures \ref{fig:network avg eps}, where we observe a positive correlation between layer size and observed discrepancy during checksum. Across all layers in the network, the observed discrepancy for DeiT-base is approximately 10x higher than that of DeiT-tiny -- going from a range of $(10^{-5}, 10^{-2})$ to $(10^{-6}, 10^{-3})$ 

\begin{figure}[t]
    \setlength{\belowcaptionskip}{-5pt}
    \setlength{\abovecaptionskip}{1mm}
    \centering
    \includegraphics[width=0.4\textwidth, clip]{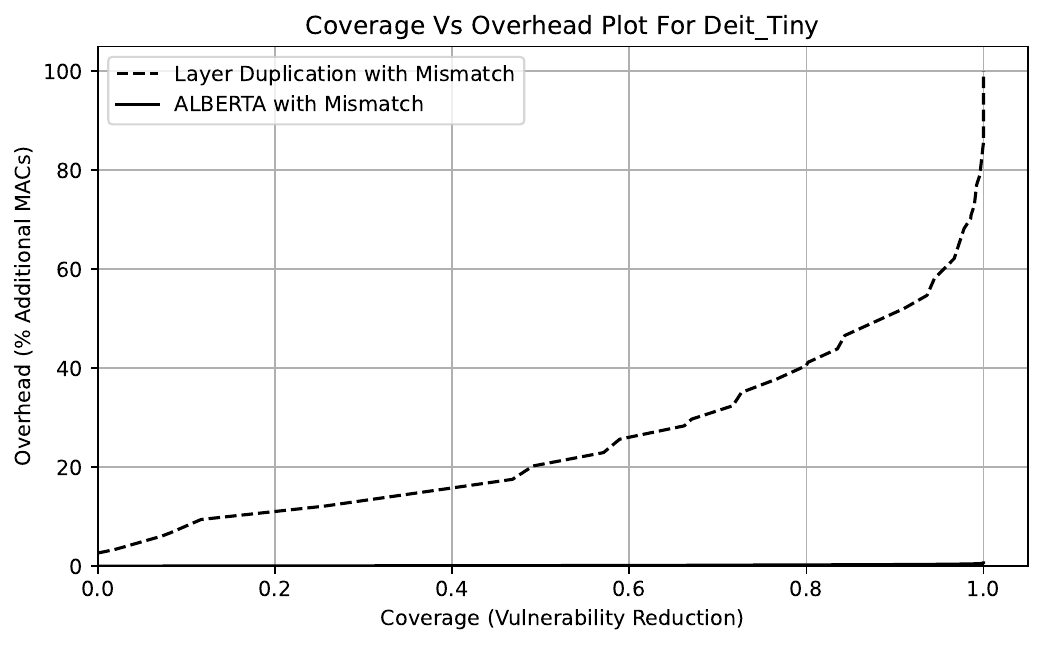}
    \vspace*{0.5mm}
    \caption{DeiT-Tiny overhead vs. coverage plot. The graph for ALBERTA lies almost flat along the x-axis and distinctly outperforms layer duplication. 
    }
    \vspace{-0.1in}
    \label{fig:deit tiny overhead}
\end{figure}
\pagebreak
\noindent\textbf{ALBERTA Coverage versus Overhead.}
As observed earlier from the layer analysis results of DeiT-base and similar models in Figure \ref{fig:v-prop}, the last layer of most transformer models is by far the most vulnerable to transient errors -- contributing to close to 50\% of all mismatches and 88\% of the total vulnerability score -- and as a result, we have designed ALBERTA such that the last layer will always be included during layer selection for protection.

Similar to the setup in Figure \ref{fig:DeiT-coverage-overhead}, the coverage \& overhead offered by protecting the last linear layer is omitted in Figure \ref{fig:deit tiny overhead}. Although it cannot be directly observed in Figure \ref{fig:deit tiny overhead}, the plots for ALBERTA -- lying flat along the x-axis -- exhibit a similar pattern to that of selective layer duplication, but is consistently over 100$\times$ more efficient. While this value depends on the dimensions of the protected layers, for most transformer models, this value is well above 100 even without considering the boosts from offline checksum generation. 

\begin{figure}[t]
    \centering
    \begin{subfigure}{0.4\textwidth}
        \includegraphics[width=\textwidth, clip]
        {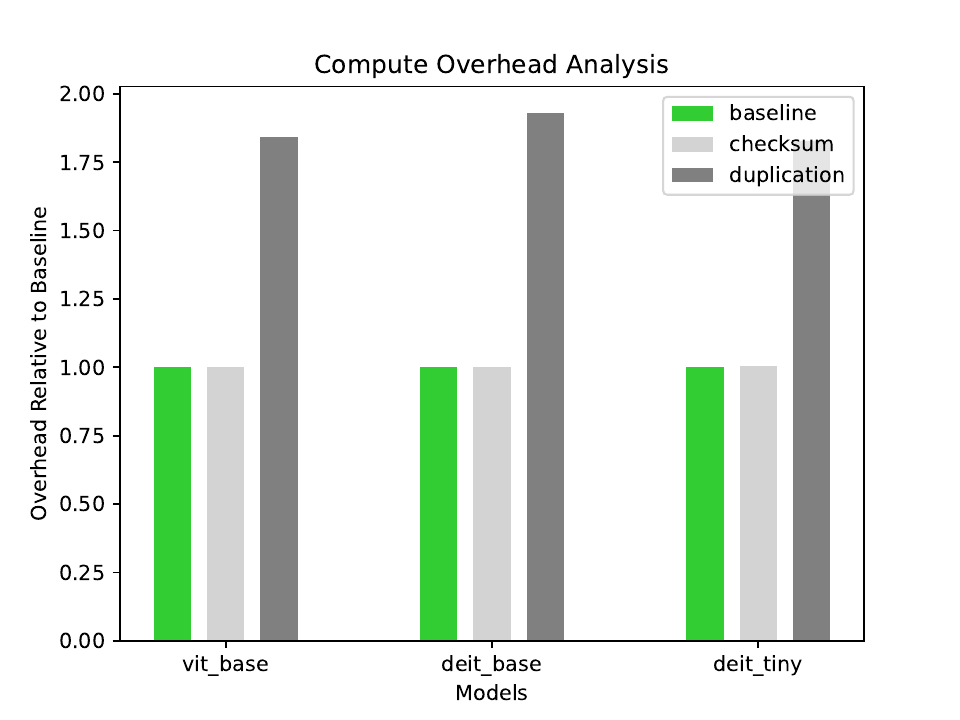}
        \vspace{-0.2in}
    \end{subfigure}    
    \begin{subfigure}{0.4\textwidth}
        \includegraphics[width=\textwidth, clip]
        {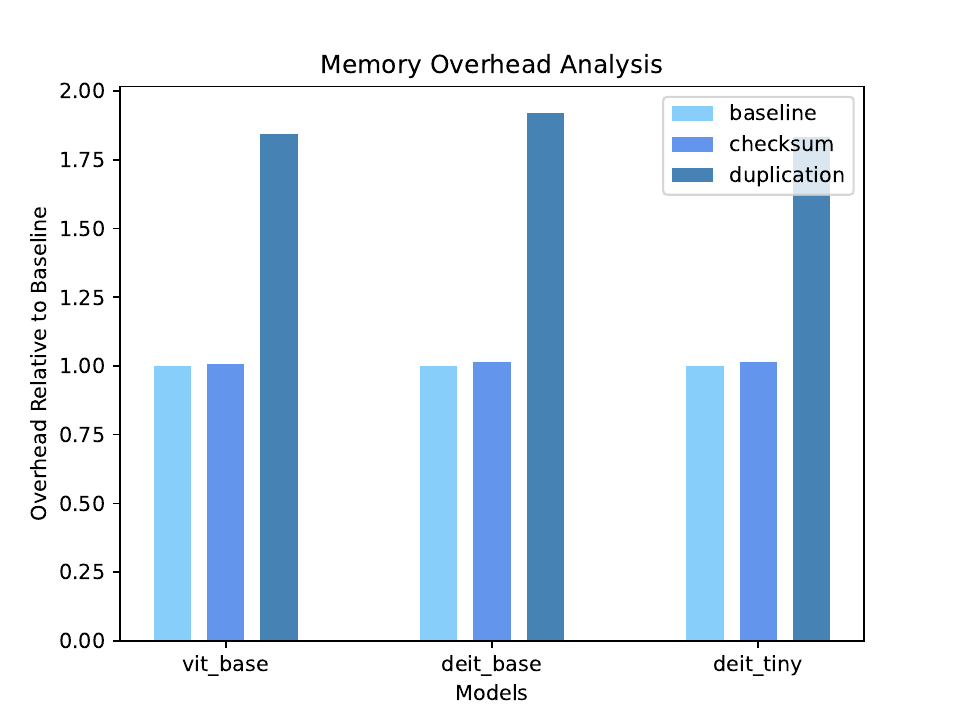}
    \end{subfigure}
    \setlength{\belowcaptionskip}{-5pt}
    \setlength{\abovecaptionskip}{-1mm}
    \caption{Network analysis of compute and memory overhead as obtained using FP64 checksums and at 99\% coverage.}
    \label{fig:network overhead analysis}
\end{figure}

In previous works such as selective duplication, serious concerns were raised due to non-negligible model performance drops that resulted from high computational and/or memory overhead. As illustrated in Figure \ref{fig:network overhead analysis}, ALBERTA marks a significant improvement from existing works where the computation overhead is less than 0.2\% across all of our target models, while the memory overhead reaches a negligible maximum of 0.01\% during checksum computation. The memory overheads are so low because only the protected layers need additional memory and that the additional memory incurred is a order of magnitude lower than the layer's original memory footprint. Specifically, the network memory overhead is computed as the sum of two components -- offline and online checksums. The offline checksum term is computed as the total memory allocated to storing the column checksum of weight matrices that are computed prior to inference \ref{fig:abed checksum overview}. The online checksum term, on the other hand, is computed as the per-layer maximum of memory allocated to input checksum vector + memory allocated to computing the output checksum during verification. In cases where kernel fusion is possible, the overhead can be significantly lower because the output checksum computation will be fused with the epilogue.

\noindent\textbf{ALBERTA Performance Analysis.}
Furthermore, assessments using \textit{torch.cuda.max\_memory\_allocated()} indicate that our target models' maximum memory usage during inference is \textgreater30\% less than that of training. Given the efficient and accelerated training support on modern GPUs, models running ALBERTA for error resilience would see little to no performance impact even in resource constrained systems. 

\begin{figure}[t]
    \setlength{\belowcaptionskip}{-5pt}
    \setlength{\abovecaptionskip}{1mm}
    \centering
    \vspace{-0.15in}
    \includegraphics[width=0.45\textwidth, clip]{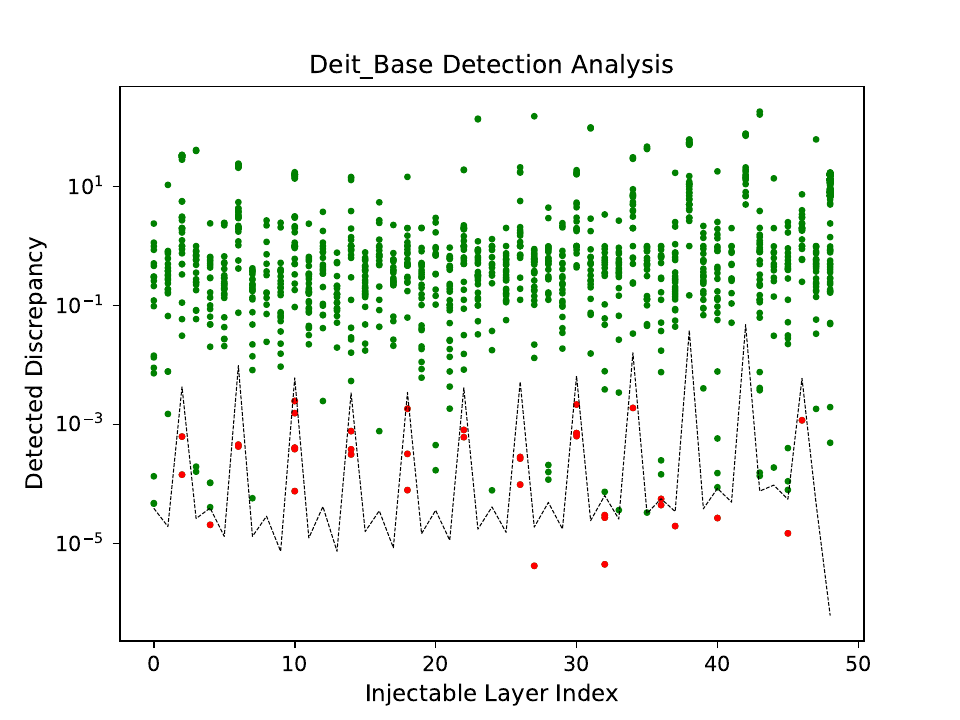}
    \caption{DeiT-base detection analysis plot. Green points represent true positive errors that are sucessfully detected, red points represent the false negatives that the detection missed, and the black line plots the per-layer discrepancy thresholds.}
    \vspace{-0.1in}
    \label{fig:detection analysis}
\end{figure}

\noindent\textbf{Effectiveness of confidence based error detection and correction.}
In related works that focus on models with fixed data types, layers protected by checksum-based algorithms easily achieves 100\% coverage of single-bit flip errors. In floating-point models, a range-based equality check at the 99.9\% confidence level does not catch all injected errors but consistently achieves over 99\% coverage of errors that result in mismatches. Figure \ref{fig:detection analysis} shows that out of 3890 error injections that resulted in mismatches, our detection algorithm achieves a false negative rate of less than 1\%, with 31 missed errors. Please note that error detection with ALBERTA has near zero false positives thanks to the earlier analysis of per-layer numerical discrepancies in \ref{fig:abed layer eps}. By setting the per layer threshold at 99.99\% confidence level or greater, we completely eliminate false positives in our detection, as any discrepancies recorded above our threshold have less than 0.01\% chance of occurring due to numerical imprecision. In our studies, errors injections that do not cause a significant change in layer activation -- magnitude of less than 1e-5 -- are not detected by our proposed framework in layers with high error-free discrepancies, but have also rarely been detected to cause a mismatch. As a general rule, there exists a correlation between how much an error offsets the original activation value and the probability of it affecting the network output. In our case with eps of less than 1e-3, such situation becomes even more unlikely to occur. 

\vspace{0.05in}

\subsection{Answering the Research Questions}
Based on the above results, we directly answer the research questions listed in Section~\ref{sec:rq} here.

\textit{RQ1:} Vision transformers are extremely resilient to transient errors when performing image classification, with injections in several layers resulting in zero mismatches. Using the origination probability and per-layer mismatches as shown in Figures 5 and 6, we compute the probability that a transient error results in an SDC for DeiT-base and DeiT-tiny to be 0.98\% and 2.49\%, respectively. While a fault in a DeiT-tiny's operation is more likely to cause an SDC, it has 13.9$\times$ fewer operations than DeiT-base.

\textit{RQ2:} Using mismatch as the ground truth vulnerability metric, we find that the prediction head of vision transformers is most vulnerable to transient errors, while many layers across the middle of the network are highly resilient. We observe a positive correlation between network size and layer level resilience (deit-base vs. deit-tiny), and that training techniques such as distillation can have a significant impact in network resilience (deit-base vs. vit-base).

\textit{RQ3:} Existing modular redundancy techniques provides good protection but at the cost of prohibitively high overheads. Despite the challenges, ALBERTA achieves over 99\% coverage for transient errors that result in a mismatch at less than 0.2\% computation overhead and on average \textless0.01\% memory overhead for vision transformer. We introduce several optimizations in the future work section that can further  reduce the memory overhead.

\textit{RQ4:} We design a reliable checksum-based error detection mechanism for floating-point networks by modeling the numerical discrepancies that occur during error-free inferences and creating per layer confidence thresholds that can be used for error detection and correction. 

\section{Discussion}

{\bf Duplication vs.\ ALBERTA:} Using mismatch as the ground truth vulnerability metric, performing selective layer duplication for DeiT-base provides 90\% coverage at around 40\% computation overhead and 92\% memory overhead. ALBERTA reduces the computation overhead to approximately 0.2\% and the memory overhead to less than 0.01\%, while also increasing the overall coverage to 99\%. By using replay as the self-correction mechanism for transient hardware faults, we are able to resolve all erroneous detection with an overhead of less than 2\% -- making the framework itself resilient to future architecture and library changes that may affect the false positive rate for error detection. \\ 

{\bf Optimizations:} While the compute and data transfer overheads are small, 
the checksum computations can be memory intensive. Several optimizations can be employed to reduce (and eliminate) this overhead. (1) L2 cache eviction policy hints (e.g., evict-last hint) available in the latest GPUs starting Ampere~\cite{cuda-l2} can be provided to keep the inputs (and outputs) persistent in L2 to minimize DRAM traffic for checksum generation. (2) Merge the checksum generation with the layer that produces the tensor. This way the tensor need not be fetched twice, eliminating extra memory traffic originating from checksum generation. (3) Lastly, hardware support to perform near L2 cache (or memory) reduction will eliminate data fetching from L2 (or memory) to streaming multiprocessors and pollute local caches with read-once data, and speed up checksum generation.

{\bf Other fault types:} Methods proposed in ALBERTA can be used to guard against SDCs from other fault sources, e.g., permanent and intermittent hardware faults, and potentially software bugs. There is increased interest in the industry to address SDCs caused by faults in processors deployed in datacenters, where the source is largely attributed to manufacturing defects and aging related faults~\cite{AlibabaSDC, MetaSDC}. While the checksum-based error detection approach will be effective in detecting such faults, the following need to change -- selection strategy to identify what to protect, a diagnosis procedure to root cause the fault type (between permanent and transient), and the recovery mechanism needs to relaunch the work on a different node to maintain forward progress.

\section{Conclusion}
\label{sec:conclusions}

We investigate the error resilience of the transformer architecture, specifically for the DeiT-base and DeiT-tiny implementations. Our investigations show that the embedding layer is most susceptible to error origination due to its large size, while the final prediction head is most vulnerable to error propagation due to its proximity to network output. Moreover, comparisons between the results for DeiT-base and DeiT-tiny serve as evidence that a larger network size contributes to lower average error propagation probability in transformers, while also highlighting the significant jump in layer resilience inside DeiT-base.

\pagebreak

\bibliographystyle{plain}

\bibliography{Bibliography}

\end{document}